\documentclass[prb,amssymb,amsmath,twocolumn]{revtex4}
\usepackage{bm}
\usepackage{graphicx}
\usepackage{color}

\begin{document}

\title{Nonlocal correlations of the local density of states in disordered quantum Hall systems}

\author{Thierry Champel}
\affiliation{Laboratoire de Physique et Mod\'{e}lisation des Milieux
Condens\'{e}s, CNRS and Universit\'{e} Joseph Fourier-Grenoble 1, B.P. 166, 
25 Avenue des Martyrs, 38042 Grenoble Cedex 9, France}

\author{Serge Florens}
\affiliation{Institut N\'{e}el, CNRS and Universit\'{e} Joseph Fourier, B.P.
166, 25 Avenue des Martyrs, 38042 Grenoble Cedex 9, France}

\author{M. E. Raikh}
\affiliation{Department of Physics, University of Utah, Salt Lake City, Utah 84112, USA}

\date{\today}

\begin{abstract}
Motivated by recent high-resolution scanning tunneling microscopy (STM) experiments
in the quantum Hall regime both on massive two-dimensional electron gas and 
on graphene, we consider theoretically the disorder averaged {\it nonlocal correlations} 
of the local density of states (LDoS) for electrons moving in a smooth disordered potential 
in the presence of a high magnetic field. The intersection of two quantum cyclotron rings 
around the two different positions of the STM tip, correlated by the local disorder, provides 
peaks in the spatial dispersion of the LDoS-LDoS correlations when the intertip distance 
matches the sum of the two quantum Larmor radii. The energy dependence displays also complex 
behavior: for the local LDoS-LDoS average ({\it i.e.} at coinciding tip positions), sharp 
positive correlations are obtained for tip voltages near Landau level, and weak 
anticorrelations otherwise.
\end{abstract}


\maketitle

\section{Introduction}

Quantum Hall systems offer a surprising dichotomy between very
universal macroscopic properties, such as the near perfect quantization
of the Hall conductance, and sample-dependent physics dominated
by local imperfections, as recently observed in several local scanning tunneling
spectroscopy (STS) experiments both on massive two-dimensional electron gas~\cite{Hashi2008} 
and on graphene~\cite{Miller2009,Miller2010}. It is, however, known that some degree of 
universality can be recovered by performing sample (or disorder) average
of local quantities, and indeed theoretical predictions for the averaged
STS local density of states (LDoS) lead to Gaussian behavior near the Landau
levels, with an energy width and a lineshape that depend on the width and correlation 
length of the disorder distribution, respectively~\cite{Wegner,Champel2010}.
Because the information extracted from transport experiments is limited,
correlations of current ({\it i.e.}, noise measurements) have been previously 
examined~\cite{Glattli}, successfully demonstrating the existence of fractionally 
charged quasiparticles for the fractional quantum Hall effect. 
The question we wish to raise here is the nature of the disorder averaged {\it nonlocal} 
correlations of {\it local} physical quantities (such as the LDoS) in the
quantum Hall regime. Such study can, in principle, be experimentally achieved
by sampling large spatial areas of the sample surface using the displacements of the STM 
tip and correlating the measured LDoS at two different tip positions (and
possibly two different tip voltages). The possibility to probe the LDoS at
different spatial locations in STM experiments offers new perspectives in comparison 
with previous experimental studies of fluctuations of the LDoS at a fixed
position, using resonant tunneling through a localized impurity 
state~\cite{Schmidt,Holder,Jouault,Konemann}.

More explicitly, from the LDoS $\rho({\bf r},\omega)$, 
which depends on tip position ${\bf r}$ and voltage $\omega$ (although we keep
the electron charge $e=-|e|$ and Planck's constant $\hbar$ in what
follows, we assume that voltage, energy and frequency are loosely identified
with each other), we define the nonlocal disorder averaged LDoS-LDoS correlations
\begin{eqnarray}
\chi(r,\omega_1,\omega_2) \equiv
\left<\rho({\bf r}_1,\omega_1) \rho({\bf r}_2,\omega_2)\right> 
- \left<\rho({\bf r}_1,\omega_1)\right> \left<\rho({\bf r}_2,\omega_2)\right> 
\nonumber \hspace*{-0.5cm} \\ 
\label{chi}
\end{eqnarray}
as the centered two-point correlation function of the LDoS (here 
$r=|{\bf r}_1-{\bf r}_2|$). Clearly this is a complicated object that 
depends on two tip voltages, but only on the distance between the two positions 
of the STM tip because of translation invariance and spatial isotropy after 
averaging.

\begin{figure}
\includegraphics[scale=0.7]{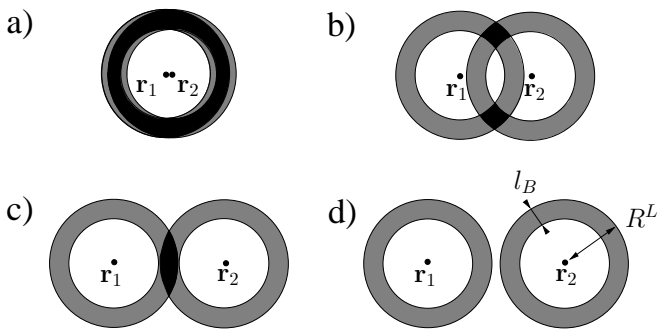}
\caption{
Geometric interpretation of correlations of the LDoS in terms of overlap of
quantum cyclotron rings of width $l_B$. Panel a) taken for close intertip distance 
$|{\bf r}_1-{\bf r}_2|\ll 2 R^L$ presents a nearly maximal overlap area of the order
$l_B R^L$ (with $R^L$ the Larmor radius), hence maximal correlations. Panel b) taken in
the intermediate situation $|{\bf r}_1-{\bf r}_2| \simeq R^L$ has a small
overlap area $l_B^2$, thus relatively weak correlations. Panel c) taken at tangent
cyclotron orbits $|{\bf r}_1-{\bf r}_2|\simeq 2 R^L$ has a small increased overlap area
of the order $l_B^{3/2} (R^L)^{1/2}$ (for $R^L>l_B$), so that a moderate peak 
as a function of intertip distance occurs. Finally panel d) presents the situation 
of non-overlaping cyclotron orbits at $|{\bf r}_1-{\bf r}_2|\gg 2 R^L$, hence 
exponentially suppressed LDoS-LDoS correlations.}
\label{ring}
\end{figure}

Before turning to detailed calculations, we wish to give some general physical 
interpretation of this physical quantity. The basic idea is that important
correlations are obtained whenever the two quantum cyclotron rings (associated to 
circular wave functions in a perpendicular magnetic field $B$) have a large 
spatial overlap (disorder plays, however, a crucial role in correlating locally 
the states, as we will see later on). Focusing first on the spatial dependence of 
$\chi(r,\omega_1,\omega_2)$
for equal tip voltages, one readily understands that the intersection of the 
quantum rings (with width of the order of magnetic length $l_B=\sqrt{\hbar c/|e|B}$) 
around the two tip positions provides an increased area of intersection 
[from $l_B^2$ to $l_B^{3/2} (R^L)^{1/2}$] when the distance between the points is close 
to the sum of the Larmor radii $R^L$ (see Fig~\ref{ring} and caption for
details). This already suggests that a peak should occur in the LDoS-LDoS correlations 
for this particular distance. 
We note that this effect has no classical analog, as quantum cyclotron rings 
collapse into circular cyclotron orbits at vanishing $l_B$ (in fact, we will see
that a kink instead of a peak occcurs in the classical limit).
The energy dependence (given by the two tip voltages) can be also easily 
inferred, for instance, at coinciding tip positions. When both tip energies precisely
match the Landau levels, maximal overlap of the whole quantum cyclotron rings occur, 
leading to sharp and large positive correlations. However, detuning the two tip
energies will probe correlations between different circular wave functions, and
destroy the correlations. In that case, the square averaged term in Eq.~(\ref{chi})
will dominate the averaged square one, leading then to weak and {\it negative} correlations.

The paper is organized as follows. Section~\ref{SecLDoS} recalls the form
of the local density of states (for fixed but smooth disorder) in quantum
Hall systems~\cite{Champel2009}. We consider, in turn, the disorder averaged LDoS 
in Sec.~\ref{DisAvLDoS}, allowing us to present the technique on a simple and more usual 
quantity, and to test the approximation scheme. Then Sec.~\ref{DisAvCor} presents
the derivation of the disorder averaged LDoS-LDoS correlations, which are
finally discussed in great detail in Sec.~\ref{Discussion}.

\section{Local density of states at high magnetic field} 
\label{SecLDoS}

We now present our theoretical analysis of this problem, which can be carried 
out fully analytically for smooth disorder in the high magnetic field regime,
and which confirms the above argumentation. The basic model is a 
single-particle Hamiltonian for an electron confined in two dimensions
in the presence of both a perpendicular magnetic field ${\bf B}$ and an 
arbitrary potential energy $V({\bf r})$,
\begin{equation}
H=\frac{1}{2 m^{\ast}} \left(-i \hbar {\bm \nabla}_{{\bf r}}-\frac{e}{c}{\bf
A}({\bf r}) \right)^{2}+V({\bf r}),
\label{Ham}
\end{equation}
with the vector potential ${\bf A}$ such that ${\bm \nabla} \times {\bf A}={\bf
B}=B \hat{{\bf z}}$, and $m^{\ast}$ the electron effective mass [here ${\bf r}=(x,y)$ 
is the position of the electron in the plane].
We do not consider the case of graphene here, which can be easily extended following 
our previous results~\cite{Champel2010}, although this system is quite relevant
experimentally for the considerations of the present work.

The starting point is the realistic assumption of large cyclotron frequency (compared 
to local amplitude fluctuations of the disordered potential) at large magnetic
field, so that Landau level mixing can be disregarded and the guiding center 
coordinate ${\bf R}$ follows a quantum motion along weakly curved equipotential lines. 
In that case the guiding center Green's function obeys a single pole structure for each 
Landau level $n$ as found in Refs. \onlinecite{Champel2009} and \onlinecite{Champel2010} [this generalizes 
the results of Ref. \onlinecite{Raikh1995} to arbitrary Landau levels]:
\begin{eqnarray}
G_{n}({\bf R})&=& \frac{1}
{ \omega-E_n-V_n({\bf R})+i0^{+} }
 \label{VortexDensity}
\end{eqnarray}
where $E_n=\hbar \omega_c(n+1/2)$ are the Landau level energies and 
$\omega_c=|e| B/(m^\ast c)$ is the cyclotron frequency. 
One important quantity above is the effective potential $V_n({\bf R})$ that results 
from averaging the bare disorder potential $V({\bf R})$ along the quantum cyclotron motion
\begin{eqnarray}
V_{n}({\bf R}) =
\int d^2 {\bm \eta} \;
F_n({\bf R}-{\bm \eta}) V({\bm \eta}) .
\label{tildev}
\end{eqnarray}
The kernel $F_n({\bf R})$ here is given by the following expression \cite{Champel2009}:
\begin{eqnarray}
 F_{n}({\bf R}) =
\frac{1}{\pi n! l_{B}^{2}}
 \frac{\partial^{n}}{\partial s^{n}}
\left. \frac{e^{-A_{s} {\bf R}^{2}/l_{B}^{2}}}{1+s} \right|_{s=0}
\label{Kn}
\end{eqnarray}
with $A_{s}=(1-s)/(1+s)$, and can also be written in an equivalent form \cite{note1}
\begin{eqnarray}
 F_{n}({\bf R}) = \frac{(-1)^{n}}{\pi l_{B}^{2}} 
L_{n}\left(\frac{2 {\bf R}^{2}}{l_{B}^{2}} \right) e^{-{\bf R}^{2}/l_{B}^{2}},
\label{Kneq}
\end{eqnarray}
where $L_{n}(z)$ is the Laguerre polynomial of degree $n$. This kernel is
also known as the form factor in the literature on quantum Hall
effects~\cite{form}.
Expression (\ref{Kn}) turns out to be very useful for the study of the first
several Landau levels, while
expression (\ref{Kneq}) is more suited for the consideration of high Landau
levels ($n \gg 1$).  Apart from the case $n=0$, we note that the kernel
$F_n({\bf R})$ is {\it not} a positive-definite function, and cannot be
interpreted as a wave function probability density. Instead, it rather 
corresponds to a Wigner distribution because the physical space of the guiding center 
coordinates ${\bf R}=(X,Y)$ is, in fact, associated to a pair of conjugate variables owing to
the commutation relation $[\hat{X},\hat{Y}]=il_B^2$ in the operatorial language. 
However, integration of the kernel $F_n({\bf R})$ over an arbitrary line provides a 
translationally invariant Landau states probability density.
In the classical limit $n \to \infty$, while keeping the (Larmor) cyclotron
radius $R^L_{n}=\sqrt{2n}l_B$ finite (hence for $l_B\rightarrow 0$), one
gets \cite{note2} $F_n({\bf R}) \simeq \frac{1}{2\pi R^L_{n}} 
\delta(|{\bf R}|-R^L_{n})$,
so that the effective potential Eq.~(\ref{tildev}) corresponds to an
average over the classical cyclotron orbit, as previously shown in
Ref.~\onlinecite{Raikh1993}. For a nonzero $l_B$, $F_n({\bf R})$
is an oscillating function that shows a sharp peak of width $l_B$ centered
around $|{\bf R}|\simeq R^L_{n}$. This quantity will be a crucial ingredient later
on for the mathematical identification of the quantum cyclotron rings.

Finally, the LDoS is readily connected~\cite{Champel2009,Champel2010} 
to the guiding center Green's function given in Eq.~(\ref{VortexDensity})
\begin{eqnarray}
\rho({\bf r},\omega) =  \int
\frac{d^{2}{\bf R}}{2 \pi l_{B}^{2}}
\sum_{n=0}^{+ \infty}
F_{n}({\bf R}-{\bf r}) \frac{-1}{\pi} \mathrm{Im}\, G_n({\bf R},\omega) .
\label{ldos}
\end{eqnarray}
Here one did not include the overall spin degeneracy. In practice, the
signal measured by local scanning tunneling spectroscopy is directly related 
to the local density of states through an energy convolution with the derivative
of the Fermi-Dirac distribution and the experimental resolution window.
Apart from the condition of high cyclotron energy, the present calculation is valid
for smooth potentials at the scale of $l_B$. Such a restriction on the spatial
variations of the potential is, in fact, not very drastic because the above results
are exact for arbitrary one-dimensional (1D) potentials~\cite{Champel2009} (even very rough ones).
For the realistic two-dimensional (2D) potential the validity of the calculation is controlled by the small 
energy scale associated typically to potential curvature~\cite{Champel2009}
\begin{equation}
E_\mathrm{curvature} = \frac{l_B^2}{2}\sqrt{|\partial_{XX}V_n\partial_{YY}V_n-
(\partial_{XY}V_n)^2|} .
\end{equation}
A more precise evaluation of the degree of reliability of our approximation scheme
for rough potentials will be given at the end of the next section. Before 
considering the correlations of the LDoS, we compute now in some detail the 
disorder averaged LDoS itself.

\section{Disorder averaged LDoS at high magnetic field}
\label{DisAvLDoS}

We first examine the disorder averaged LDoS, obtained experimentally
by spatially sampling the STS current over a single STM tip position.
Theoretically, the averaging procedure of expression~(\ref{ldos}) will be carried 
through an isotropic distribution function $\tilde{C}(q)$ in Fourier space 
(here $q=|{\bf q}|$) that describes the spatial correlations of disorder
\begin{eqnarray}
\label{average}
\left<V({\bf R}_1) V({\bf R}_2)\right> &=& C({\bf R}_1-{\bf R}_2)  \\
&=& \int \frac{d^2 {\bf q}}{(2\pi)^2} \tilde{C}(q) \, e^{-i{\bf q}\cdot({\bf R}_1-{\bf R}_2)}.
\nonumber
\end{eqnarray}
Typically one can take $C({\bf R})=v^2 e^{-|{\bf R}|^2/\xi^2}$, defining the 
correlation length $\xi$ and the root mean square value 
$v=\sqrt{\left<V({\bf R})^2\right>}$ of the bare disorder distribution.
In that case $\tilde{C}(q)=\pi v^2\xi^2 e^{-\xi^2 q^2/4}$.
Averaging of the LDoS (\ref{ldos}) then simply follows from exponentiating the
single pole in Eq.~(\ref{VortexDensity}) by going in the time domain, and performing Gaussian
integration over all possible disorder realizations 
\begin{eqnarray}
\label{pasdenom}
\left<\rho({\bf r},\omega)\right> &=&
\int \!\!\! \frac{dt}{2\pi} 
\int \!\!\!  \frac{d^{2}{\bf R}}{2 \pi l_{B}^{2}}
\sum_{n=0}^{+ \infty} 
F_{n}({\bf R}-{\bf r}) 
\int \!\!\! \mathcal{D}V \, \\
\nonumber
&& \hspace{-1.2cm}
\times \exp\left\{
i[\omega-E_n-V_n({\bf R})] t
-\frac{1}{2} \int \frac{d^2{\bf q}}{(2\pi)^2} 
\frac{|\tilde{V}({\bf q})|^2}{\tilde{C}(q)}  \right\}.
\end{eqnarray} 
The effective potential Eq.~(\ref{tildev}) is given in Fourier
space by $\tilde{V}_n({\bf q}) = \tilde{F}_n(q) \tilde{V}({\bf q})$, where $\tilde{F}_n(q)$
is the Fourier transform of kernel~(\ref{Kn}), which is easily shown
to obey
\begin{eqnarray}
\nonumber
 \tilde{F}_{n}(q) &=&  
\int d^2{\bf r} \, e^{i{\bf q}.{\bf r}} F_n({\bf r}) 
= \frac{1}{n!}
 \frac{\partial^{n}}{\partial s^{n}}
\left. \frac{e^{-l_B^2 q^{2}/(4 A_s)}}{(1+s)A_s} \right|_{s=0}\\
&=& (-1)^n \pi l_B^2 \, F_n(l_B^2 {\bf q}/2).
\label{KnFourier}
\end{eqnarray}
One can then readily perform the functional integral over the disorder 
realizations in Eq.~(\ref{pasdenom}):
\begin{eqnarray}
\nonumber
\left<\rho({\bf r},\omega)\right> &=&
\int \!\!\! \frac{dt}{2\pi}
\int \!\!\!
\frac{d^{2}{\bf R}}{2 \pi l_{B}^{2}}
\sum_{n=0}^{+ \infty} 
e^{i(\omega-E_n) t}
F_{n}({\bf R}-{\bf r}) \\
&& \times \exp \left\{ -\frac{1}{4} t^2\, \Gamma_n^2 \right\}
\label{rhoint}
\end{eqnarray} 
where the energy width $\Gamma_n$ is given by the relation
\begin{eqnarray}
\Gamma_n^2 = 2 \int \!\!\! \frac{d^2 {\bf q}}{(2\pi)^2} 
\tilde{C}(q) \left| \tilde{F}_n(q) \right|^2 .
\label{width}
\end{eqnarray}
This result was obtained initially in Ref. \onlinecite{Champel2010}
for the case of graphene.

Expression~(\ref{rhoint}) for the averaged density of states (DoS) is 
obviously ${\bf r}$ independent, so that the ${\bf R}$ integral can be 
carried using the normalization condition $\int d^2{\bf R}\, F_n({\bf R})=1$. 
The remaining time integral 
gives the final result
\begin{eqnarray}
\label{averageDos}
\hspace{-0.3cm}
\left<\rho({\bf r},\omega)\right> =
\frac{1}{2\pi l_B^2}
\sum_{n=0}^{+\infty}
\frac{1}{\sqrt{\pi} \Gamma_{n}}
\exp\left\{-\left(\frac{\omega- E_n}
{\Gamma_n}\right)^2\right\} ,
\end{eqnarray}
which takes the expected Gaussian lineshape.

\begin{figure}
\includegraphics{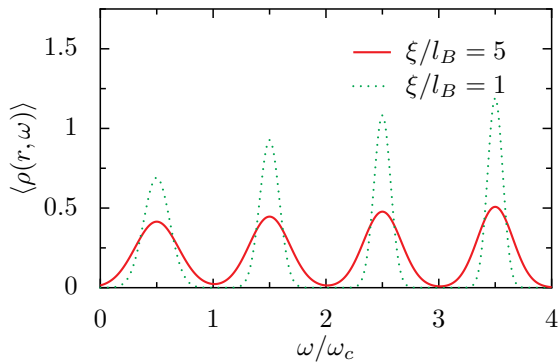}
\caption{(color online) Disorder averaged density of states [in units of $(2\pi
l_B^2 v)^{-1}$] as a function of energy for two values of the correlation length 
of the disorder distribution ($\xi/l_B=5$ and $\xi/l_B=1$) spanning the first 
four Landau levels.}
\label{dos}
\end{figure}

The renormalized disorder width $\Gamma_n$ given in Eq.~(\ref{width}) can be 
analyzed  \cite{note2} in the classical limit $n\rightarrow+\infty$, 
keeping the cyclotron radius $R^L_{n}=\sqrt{2n}l_B$ fixed and  $l_{B} \to 0$.
In this regime, we recover results first derived in Ref.~\cite{Raikh1993} using
a completely unrelated method
\begin{eqnarray}
\Gamma_n =  \sqrt{\int \frac{q dq}{2\pi} 
2 \tilde{C}(q) [J_0(R^L_{n} q)]^2} ,
\label{newwidth}
\end{eqnarray}
with $J_0(z)$ the zeroth order Bessel function. 
For very large classical orbits such that $R_n^L \gg \xi$,
the large-argument asymptotics of the zeroth order 
Bessel function $J_0(z)\simeq\sqrt{2/(\pi z)}\cos(z-\pi/4)$ can be used, so that
\begin{eqnarray}
\Gamma_n \simeq  \sqrt{\frac{\int dq\, 2 \tilde{C}(q)}{\pi^2 R^L_{n}}}
 \propto \frac{1}{n^{1/4}},
\label{approxwidth}
\end{eqnarray}
showing a decrease of the Landau level energy width with increasing index $n$.
We note that expression~(\ref{width}) is more general than the classical result (\ref{newwidth}) 
because it incorporates wave function spreads on the scale $l_B$, a purely quantum lengthscale 
which has completely disappeared in the classical limit.
In all cases (classical or quantum), the general trend is that the cyclotron motion 
averages out the local potential at increasing radius $R^L_{n}$, so that the energy width of 
the sample averaged DoS decreases with $n$. This effect is clearly 
seen~\cite{Hashi2008,Miller2009,Miller2010} from the experimentally measured spatial dispersion 
of the LDoS, which shows a rapid narrowing for higher Landau levels.
However, in the opposite limit of very smooth disorder $\xi\gg R^L_{n}$, this averaging by
the cyclotron orbits becomes less efficient, and the energy width $\Gamma_n$
depends very weakly on the Landau level index $n$. Both regimes are presented in
Fig.~\ref{dos} showing the energy-dependent disorder averaged density of states
for two values of $\xi/l_B$.

\begin{figure}
\includegraphics{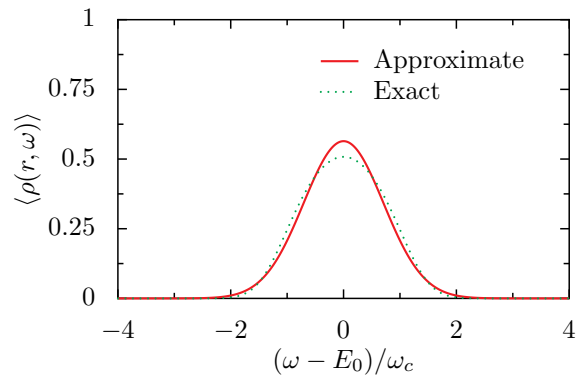}
\caption{(color online) Comparison of the disorder averaged density of states
[in units of $(2\pi l_B^2 \Gamma_0)^{-1}$] in
the Landau level $n=0$ between Wegner's exact formula and our leading order
gradient approximation for the extreme case of $\delta$-correlated disorder
($\xi/l_B=0$). Rapid convergence of the approximation scheme is expected as soon
as $\xi\gtrsim l_B$.}
\label{Wegner}
\end{figure}

We end up by commenting on the reliability of our calculations for realistic
disorders, and possibly rough ones at the scale of $l_B$. From the construction of the 
quantum guiding center theory~\cite{Raikh1995,Champel2009} as a systematic gradient expansion, the
lowest order calculations used here become exact for smooth disorder, namely
$\xi\gg l_B$. In the opposite limit $\xi\ll l_B$, the approximation scheme does not 
fully break down {\it per se}, thanks to the wave function averaging on the scale 
$l_B$ performed within the effective potential~(\ref{tildev}). This implies that 
the effective potential varies on the scale $\sqrt{\xi^2+l_B^2}$ and remains smooth even
for very rough bare potential ($\xi\ll l_B$). In this regime, the theory simply misses 
then the small parameter $l_B^2/(\xi^2+l_B^2)$ of the case $\xi\gg l_B$, and its adequation 
becomes purely a quantitative matter.
Interestingly, we are able to assess its validity thanks to Wegner's exact
solution~\cite{Wegner} for the disorder averaged density of states in the Landau level $n=0$, 
calculated for $\delta$-correlated disorder $\left< V({\bf R}) V({\bf 0})
\right> = w\, \delta^{(2)}({\bf R})$
\begin{eqnarray}
\nonumber
\left<\rho({\bf r},\omega)\right>_\mathrm{exact} = 
\frac{1}{2\pi l_B^2} \frac{2^{3/2}}{\pi^{3/2} \Gamma_{0}}
\frac{\exp[2(\omega-E_0)^2/\Gamma_0^2]}
{1+\{\mathrm{Erfi}[\sqrt{2}(\omega-E_0)/\Gamma_0)]\}^2}
\hspace*{-0.6cm} \\
\label{rhoExact}
\end{eqnarray}
In the previous formula, $\mathrm{Erfi}(x) = 2
\pi^{-1/2} \int_0^x \mathrm{d}u \, e^{u^2}$ is the so-called complex error function
and $\Gamma_0^2=w/(\pi l_B^2)$.
In our calculation, the $\delta$-correlated potential is obtained from the
$\xi\rightarrow0$ limit in Eq. (\ref{average}), which corresponds to
the most stringent limit to test our approximation scheme. 
The comparison (derived for the same microscopic disorder parameters)
between the Gaussian expression~(\ref{averageDos}) in the lowest
Landau level $n=0$, using the linewidth~(\ref{width}), with Wegner's exact
formula~\cite{Wegner}, written in Eq.~(\ref{rhoExact}), is given in Fig.~\ref{Wegner}.
Although neither the peak value nor the tails (not shown) are exactly reproduced 
within our approximation, the result is close to the exact answer, and this is 
surprising because we are far from the naive domain of validity of the theory. 

This comparison gives some evidence that both the disorder averaged LDoS and its 
nonlocal correlations (to be calculated now), will be quantitatively obtained by 
a gradient expansion in the disorder for realistic cases where $\xi\gtrsim l_B$ 
(granted by the high magnetic field limit). Moreover, we stress that Wegner's method 
does not apply to obtain two-particle averages, and cannot be used to calculate the 
LDoS-LDoS correlations, showing the greater generality of our approach.

\section{Disorder averaged LDoS-LDoS correlations at high magnetic field}
\label{DisAvCor}

We are now ready to calculate the two-point correlations of the LDoS, taken at 
two different tip positions ${\bf r}_1, {\bf r}_2$ and for two different 
energies $\omega_1$, $\omega_2$. The starting point is the expression
\begin{eqnarray}
\nonumber
\!\!\left<\rho({\bf r}_1,\omega_1) \rho({\bf r}_2,\omega_2)\right> \!\!\! &=& \!\!\!
\int \!\!\! \frac{dt_1}{2\pi}\!\! 
\int \!\!\! \frac{dt_2}{2\pi}\!\!
\int \!\!\!  \frac{d^{2}{\bf R}_1}{2 \pi l_{B}^{2}}\!\!
\int \!\!\!  \frac{d^{2}{\bf R}_2}{2 \pi l_{B}^{2}}\!\!
\sum_{n_1=0}^{+ \infty}
\sum_{n_2=0}^{+ \infty}
\\
\nonumber
&& \hspace{-3.5cm} \times
F_{n_1}({\bf R}_1-{\bf r}_1) 
F_{n_2}({\bf R}_2-{\bf r}_2) 
e^{i(\omega_1-E_{n_1}) t_1 +
i(\omega_2-E_{n_2}) t_2}\\
&& \hspace{-3.5cm} \times
A({\bf R}_1,n_1,t_1;{\bf R}_2,n_2,t_2)
\label{rhorho}
\end{eqnarray}
where the following disorder average must be performed:
\begin{eqnarray}
\nonumber
A({\bf R}_1,n_1,t_2; {\bf R}_2,n_2,t_2) \!\!&=&\!\!
\int \!\!\! \mathcal{D}V \,
e^{-i[V_{n_1}({\bf R}_1) t_1
+V_{n_2}({\bf R}_2) t_2]}\\
\nonumber
&& \hspace{-0.7cm} \times
\exp\left\{
-\frac{1}{2} \int \frac{d^2{\bf q}}{(2\pi)^2} 
\frac{|\tilde{V}({\bf q})|^2}{\tilde{C}(q)}  \right\} \\
\nonumber
&&  \hspace{-4.5cm} =
\exp\left\{-\frac{1}{2}
 \int \!\!\! \frac{d^2 {\bf q}}{(2\pi)^2} 
% C(q) \left| t_1 F_{n_1}(q) e^{i {\bf q}.{\bf R}_1} 
%+ t_2 F_{n_2}(q) e^{i {\bf q}.{\bf R}_2} 
\tilde{C}(q) \Big| \sum_{i=1,2} t_i \tilde{F}_{n_i}(q) e^{i {\bf q}.{\bf R}_i} \Big|^2
\right\}\\
\nonumber
&&  \hspace{-4.5cm} =
\exp\left\{-\frac{1}{4}\left[ t_1^2 \Gamma_{n_1}^2
+ t_2^2 \Gamma_{n_2}^2 + 2 t_1 t_2 S_{n_1,n_2}({\bf R}_1-{\bf R}_2)
\right]
\right\}.
\end{eqnarray} 
The energy width $\Gamma_n$ was already defined in Eq.~(\ref{width}),
and a {\it spatially dependent} disorder correlator now appears
\begin{eqnarray}
\!\!\!
S_{n_1,n_2}({\bf R})
= 2 \int \!\!\! \frac{d^2 {\bf q}}{(2\pi)^2} 
\tilde{C}(q)  \tilde{F}_{n_1}(q) \tilde{F}_{n_2}(q) \cos({\bf q}\cdot{\bf R})
\label{newcor}.
\end{eqnarray}
Using expressions (\ref{Kn}) and (\ref{KnFourier}), we can perform exactly the
integrals in Eq. (\ref{newcor}) and find a result which can be written under the
form 
\begin{eqnarray}
\!\!\!
S_{n_1,n_2}(R)
=  \frac{2v_{0}^{2} }{ n_1!n_2!}  \frac{\partial^{n_1+n_2}}{ \partial s_{1}^{n_1} \partial s_{2}^{n_2} } 
\Bigg[ (1-s_1-s_2+s_1s_2)^{-1} \nonumber \\
\times  \frac{\xi^{2}}{\xi^2+2B_{s_1,s_2}l_{B}^{2}}  \exp\left(-\frac{R^2}{\xi^2+2 B_{s_1,s_2}l_B^2} \right)
 \Bigg]_{s_1=s_2=0},
\label{newcor2}
\end{eqnarray}
where we have introduced the short-hand notation 
\begin{equation}
B_{s_1,s_2}=\frac{1-s_1 s_2}{1-s_1-s_2+s_1 s_2}.
\end{equation}
Note that the disorder correlator $S_{n_1,n_2}(R)$ is  isotropic (i.e.,
only depends on the distance $R=|{\bf R}|$). In addition, its diagonal elements
taken at $R=0$ are related to the energy width $\Gamma_n$ via the relation
$S_{n,n}(0)=\Gamma_n^2$. For a smooth disordered potential such that $\xi \gg
l_B$, we can easily check from Eq. (\ref{newcor2}) that the functions
$S_{n_1,n_2}(R)$ are peaked at $R=0$ and decay in a Gaussian way with a
characteristic length scale given by $\xi$ for the first few Landau levels. For
higher Landau levels, the disorder correlator $S_{n_1,n_2}(R)$ spreads over a
bigger characteristic distance. For instance, for $n_1=n_2=n \gg 1$, it has an
extent of the order of $\sqrt{(R^L_{n})^2+\xi^2}$.

Shifting the space integrals in Eq. (\ref{rhorho}), we see that the quantity 
$\left<\rho({\bf r}_1,\omega_1) \rho({\bf r}_2,\omega_2)\right>$  is clearly described 
by a function of ${\bf r}={\bf r}_1-{\bf r}_2$ only.
In the following we will consider the centered two-point correlator of 
the LDoS defined in Eq.~(\ref{chi}).
Integration over the time variables in Eq.~(\ref{rhorho}) can  
be performed analytically and yields the expression for the LDoS-LDoS correlator
\begin{eqnarray}
\nonumber
\chi({\bf r},\omega_1,\omega_2) =  
\int \!\!\!  \frac{d^{2}{\bf R}_1}{2 \pi l_{B}^{2}}\!\!
\int \!\!\!  \frac{d^{2}{\bf R}_2}{2 \pi l_{B}^{2}}\!\!
\sum_{n_1=0}^{+ \infty}
\sum_{n_2=0}^{+ \infty}\!\! F_{n_1}({\bf R}_1-{\bf r}_1) 
 \\  
\times 
F_{n_2}({\bf R}_2-{\bf r}_2) \, 
\Theta_{n_1,n_2}\left(\left|{\bf R}_1-{\bf R}_2 \right|,\omega_1,\omega_2\right) \hspace*{0.6cm}
\label{rhorhofinal}
\end{eqnarray}
with
\begin{eqnarray}
\Theta_{n_1,n_2}\left(R,\omega_1,\omega_2\right) =
\frac{1}{\pi} \left\{\left[\Gamma_{n_1}^2\Gamma_{n_2}^2-[S_{n_1,n_2}(R)]^2 \right]^{-1/2} \nonumber \right.   \\
 \times
e^{- \frac{
(\omega_1-E_{n_1})^2 \Gamma_{n_2}^2 
+(\omega_2-E_{n_2})^2 \Gamma_{n_1}^2 
-2(\omega_1-E_{n_1})(\omega_2-E_{n_2}) S_{n_1,n_2}(R) }
{ \Gamma_{n_1}^2\Gamma_{n_2}^2-[S_{n_1,n_2}(R)]^2}
} \nonumber   \hspace*{-0.2cm}
\\
- 
\left[\Gamma_{n_1} \Gamma_{n_2}\right]^{-1}
e^{-\frac{(\omega_1-E_{n_1})^2}{\Gamma^{2}_{n_1}}} e^{-\frac{(\omega_2-E_{n_2})^2}{\Gamma^{2}_{n_2}}} \left. \frac{}{}
\right\}. \hspace*{0.5cm} 
 \label{Theta}
\end{eqnarray}
Although the above expressions are still too complicated to make precise
statements on the nature of the LDoS-LDoS correlations, we can already infer
some of the early predictions formulated in the Introduction. Clearly the two kernels
$F_n({\bf R-r})$ in Eq.~(\ref{rhorhofinal}) centered on the two positions of the tip
${\bf r}_1$ and ${\bf r}_2$
impose a constraint on the two guiding center coordinates ${\bf R}_1$ and ${\bf
R}_2$, which must live predominantly within cyclotron rings of extent
$R^L_{n_1}$ and $R^L_{n_2}$ with width $l_B$ (we note 
that the kernels have rather $n$ oscillations within a given radius $R^L_{n}$, so that 
there are in total $n$ different rings for a given quantum state). This formula 
thus confirms our initial expectation that the area of overlap between two cyclotron rings
dictates the behavior of the LDoS-LDoS correlations. A second interesting aspect
is that it is the additional disorder-dependent kernel $\Theta_{n_1,n_2}$
which permits nontrivial correlations. Should this complicated function of
energy and space be negligible, one would end up with vanishing correlations.

Yet, to get more quantitative understanding of the complete integral
in Eq.~(\ref{rhorhofinal}), one needs to further simplify the above expressions.
This can be achieved by introducing the change in variables 
${\bf R}'=({\bf R}_1+{\bf R}_2)/2$ and ${\bf R}={\bf R}_2-{\bf R}_1$. 
After integration over the center-of-mass position ${\bf R}'$, we get
\begin{eqnarray}
\nonumber
\chi({\bf r},\omega_1,\omega_2) =  \sum_{n_1=0}^{+ \infty}  \sum_{n_2=0}^{+ \infty}
\int \!\!\!  \frac{d^{2}{\bf R}}{2 \pi l_{B}^2}  
f_{n_1,n_2}({\bf R}+{\bf r}) \\ 
\times  \Theta_{n_1,n_2}\left(R,\omega_1,\omega_2\right)
\label{Step1}
\end{eqnarray}
with
\begin{eqnarray}
\nonumber
f_{n_{1},n_{2}}({\bf R}) = \int \!\!\!  \frac{d^{2}{\bf R}'}{2 \pi l_{B}^{2}}  
F_{n_1}\left({\bf R}'-\frac{{\bf R}}{2} \right) 
F_{n_2} \left({\bf R}'+\frac{{\bf R}}{2} \right)  \label{f} \\
=  \frac{(2 \pi l_B^2)^{-2}}{ n_1!n_2!} \left. \frac{\partial^{n_1+n_2}}{ \partial s_{1}^{n_1} 
\partial s_{2}^{n_2} } \frac{\exp\left(-  {\bf R}^2/\left[ 2 l_{B}^2  B_{s_1,s_2} 
\right] \right)}{1-s_1 s_2}  \right|_{s_1=s_2=0},
\end{eqnarray}
where we have used expression (\ref{Kn}) for the kernel $F_n({\bf R})$. 
Then, introducing the polar coordinates for the position ${\bf R}$ and
performing the angular integral in Eq. (\ref{Step1}), we arrive at the final
expression of the disorder averaged LDoS-LDoS correlations (which now
obviously is only function of the intertip distance $r=|{\bf r}|$):
\begin{eqnarray}
\nonumber
 \chi(r,\omega_1,\omega_2)  = \sum_{n_1=0}^{+ \infty}
\sum_{n_2=0}^{+ \infty} l_{B}^{-2}
\int_{0}^{+ \infty} \!\!\!\!\!\!\!  R dR \,
h_{n_1,n_2}(R,r) \\ 
\times
\Theta_{n_1,n_2}\left(R,\omega_1,\omega_2\right)
\label{Step2},
\end{eqnarray}
where 
\begin{eqnarray}
h_{n_{1},n_{2}}(R,r) =
\int_{0}^{2 \pi} \!\! \frac{d \theta }{2 \pi } \, f_{n_1,n_2}\left( \sqrt{R^2+r^2+2 rR \cos \theta}\right) \nonumber
\\
=  \frac{(2 \pi l_B^2)^{-2}}{ n_1!n_2!} \frac{\partial^{n_1+n_2}}{ \partial s_{1}^{n_1} \partial s_{2}^{n_2} }
\,\Bigg[
I_{0}\left( \frac{rR}{l_B^2 B_{s_1,s_2}} \right) \hspace*{0.6cm} \nonumber \\
  \times
 \frac{\exp\left(-  \left[ {\bf R}^2+{\bf r}^2\right] 
/\left[2 l_{B}^2 B_{s_1,s_2} \right]   \right)}{1-s_1 s_2}  \Bigg]_{s_1=s_2=0}
\hspace*{0.5cm}
\label{h}
\end{eqnarray}
with $I_{0}(x)$ the modified Bessel function of the first kind.
An alternative writing of Eq. (\ref{h}) obtained by going to the Fourier space, which turns 
out to be more suitable for the consideration of large $n_1$ and $n_2$, is as follows:
\begin{eqnarray}
 h_{n_1,n_2}(R,r)= \frac{l_B^2}{ \left( 2 \pi l_B^2 \right)^{2}}  \int_{0}^{+ \infty} 
\!\!\! \!\!\!\!\! dq \, q \tilde{F}_{n_1}(q) \tilde{F}_{n_2}(q) J_{0}(Rq) \nonumber \\
\times J_{0}(rq).
\label{halt}
\end{eqnarray}
The above expressions (\ref{Step2}) through (\ref{halt}) are exact for arbitrary 2D smooth 
disorder ({\it i.e.}, $\xi\gg l_B$).

\section{Discussion of the LDoS-LDoS correlations}
\label{Discussion}

\subsection{Spatial dependence of the correlations}

Let us now analyze the LDoS-LDoS correlations on the basis of Eq. (\ref{Step2}).
We see that the correlations result from the combination of two functions
$h_{n_1,n_2}(R,r)$ and $\Theta_{n_1,n_2}\left(R,\omega_1,\omega_2\right)$, which
are quite different in nature. 
Obviously, only the functions $\Theta_{n_1,n_2}\left(R,\omega_1,\omega_2\right)$
defined in Eq. (\ref{Theta}) contain the information about the energy
dependence.  For sufficiently widely separated Landau levels, the overlap
between two states with two given energies $\omega_1$ and $\omega_2$ vanishes
for most of the Landau level pairs due to the sharp energy Gaussians cutoff in
Eq. (\ref{Theta}), so that essentially only the one specific pair of Landau levels
$(n_1,n_2)$ that is associated to the cyclotron energies $(E_{n_1},E_{n_2})$
closest to the tip voltages $(\omega_1,\omega_2)$ yields a nonzero
contribution in the sum over Landau levels indices in Eq. (\ref{Step2}).
Furthermore, the LDoS-LDoS correlator $\chi(r,\omega_1,\omega_2)$ clearly
vanishes whenever the disorder correlator $S_{n_1,n_2}(R)$ is small compared to
$\Gamma_{n_1} \Gamma_{n_2}$ due to the exact cancellation by the square
averaged term in Eq.~(\ref{Theta}). Therefore, the main contributions to the
correlations arise when $S_{n_1,n_2}(R) \sim \Gamma_{n_1} \Gamma_{n_2}$.

In contrast, the functions $h_{n_1,n_2}(R,r)$ are independent of the
characteristic features of the disorder and have a pure geometric origin (as
already discussed, they contain information on the spatial overlap of the
quantum cyclotron rings). For $r=0$, we have the relation 
$h_{n_1,n_2}(R,0)=f_{n_1,n_2}({\bf R})$ from which
a simple physical interpretation can be easily drawn. 
Using the semiclassical physical picture of the kernel $F_n({\bf R})$ put forward previously,
we understand that nonzero contributions for $f_{n_1,n_2}({\bf R})$ in
Eq. (\ref{f}) are picked up only from the (quantum broadened on scale $l_B$)
cyclotron orbits with radii $R^L_{n_1}$ and $R^L_{n_2}$ that live around the points
$\pm {\bf R}/2$ separated by the distance $R$. 
This statement can be proved on more mathematical grounds starting from
expression (\ref{halt}).  Taking the limits $n_1$ and $n_2 \to \infty$  in Eq.
(\ref{halt}) as done in Ref. \onlinecite{note2}, we obtain an approximate formula,
which reads
\begin{eqnarray}
 h_{n_1,n_2}(R,r) \simeq  \frac{l_B^2}{ \left( 2 \pi l_B^2 \right)^{2}}  
\int_{0}^{+ \infty} \!\!\! \!\!\!\!\! dq \, q J_{0}(R^L_{n_1}q) \, J_{0}(R^L_{n_2}q) \nonumber \\
\times  J_{0}(Rq) \, J_{0}(rq).
\label{halt2}
\end{eqnarray}
For $r=0$, integral (\ref{halt2}) strictly vanishes when $R > R^L_{n_1}+R^L_{n_2}$ 
or $R < \left|R^L_{n_1} -R^L_{n_2}\right|$. In the more realistic case of finite
$n_1$ and $n_2$, this semiclassical limit shows that the resulting overlap of two quantum 
orbital motions turns out to be significant under the inequalities 
$\left|R^L_{n_1}-R^L_{n_2}\right|< R< R^L_{n_1}+R^L_{n_2}$. The functions 
$h_{n_1,n_2}(R,r)$ being symmetrical in $R$ and $r$, we understand that for $R=0$ the 
contributions to the LDoS-LDoS correlations arise when the distance $r$ between the two tip
positions is such that  $\left|R^L_{n_1}-R^L_{n_2}\right|< r< R^L_{n_1}+R^L_{n_2}$,
precisely as anticipated in the discussion of Fig. \ref{ring}.

\begin{figure}[ht]
\includegraphics{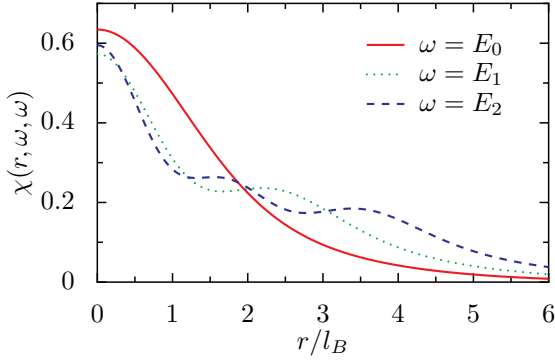}
\caption{(color online) 
LDoS-LDoS correlations at equal energies ($\omega_1=\omega_2=\omega$) as a
function of the tips distance $r$ (in units of $l_B$) for different energies
$\omega$. Here the correlations are expressed in units of $(2 \pi l_{B}^2 v)^{-2}$ 
with the parameters $\hbar \omega_c/v=5$ and $l_B/\xi=0.2$. The
spatial dependence of the correlations presents peaks whose number
corresponds to the Landau level index of the probed energy (for instance, for
$\omega=E_2$ two peaks are seen).}
\label{Fig2}
\end{figure}

The spatial dependence of the two-point correlator $\chi(r,\omega_1,\omega_2)$ 
resulting from the numerical computation of the integral
over the distance $R$ in Eq. (\ref{Step2}) is shown in Fig. \ref{Fig2} 
in the regime of widely separated Landau peaks (we have taken in
what follows $\hbar \omega_c=5 \, v$). In these two figures, we have chosen the
situation where the LDoS-LDoS correlations are maximal, that is, we have considered
that the energies $\omega_i$ (here $i=1,2$) correspond exactly to Landau level energies
$E_{n_i}$.  The case of equal energies $\omega_1=\omega_2=\omega$ is first
investigated in Fig. \ref{Fig2}. According to the previous discussion, the
dominant contributions among the different possible pairs of Landau level
indices are the diagonal ones corresponding to $n_1=n_2=n$.  For the lowest
Landau level energy ($n=0$) shown with the solid line in Fig. \ref{Fig2}, the
correlations decrease in a monotonic way as a function of the distance $r$
between the two tip positions with a characteristic decay length of the order of
the disorder correlation length $\xi$  (here all the distances are expressed in
units of the magnetic length $l_B$ and we have taken $\xi=5 \,l_B$). When the
energy of the first Landau level is probed (dotted line of Fig. \ref{Fig2}
corresponding to $\omega=E_1$), the spatial dependence of the correlations is
still decreasing with the same decay length $\xi$, but it now exhibits a mild
peak for the position $r$ close to $2 R^L_{n}= 2 \sqrt{2 n}l_B \approx 2.8 l_B$ for $n=1$, 
associated to the overlap of the quantum cyclotron rings for the $n=1$ Landau states.
In the second Landau level (at $\omega=E_2$), besides the peak close to 
$2 R^L_{n} = 4 l_B$ for $n=2$, an additional peak is clearly seen in the $r$ dependence 
of the LDoS-LDoS correlations, see dashed line in Fig. \ref{Fig2}.
This is because the wave functions for $n>1$ have $n$ zeros, hence
additional rings of high probability density. 

\begin{figure}[ht]
\includegraphics{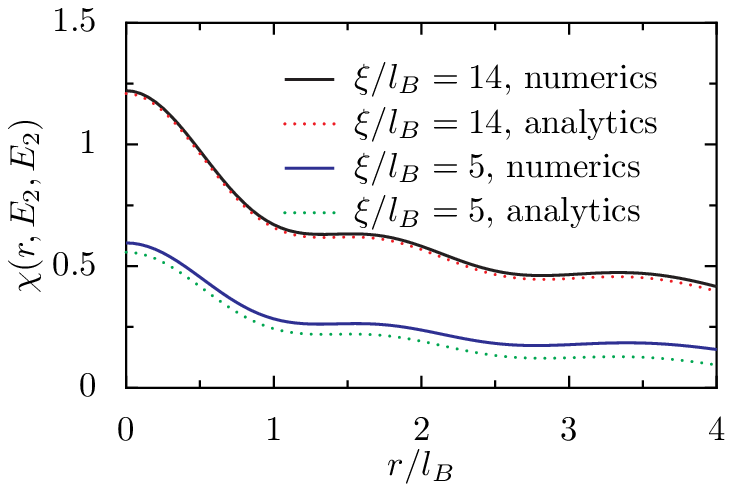}
\caption{(color online) 
LDoS-LDoS correlations at equal energies taken at the second Landau level 
($\omega_1=\omega_2=E_2$) as a function of the tips distance $r$ (in units of $l_B$) 
for two different values of the disorder correlation length $\xi/l_B=14$ and $\xi/l_B=5$, 
with a comparison of the numerical evaluation of expression~(\ref{Step2}) and the
analytical formula~(\ref{chiapprox}).}
\label{compare}
\end{figure}

This nonmonotonic spatial dependence observed for equal energies was anticipated
in the Introduction of the paper, and can be understood more quantitatively as follows. 
When the disorder potential is smooth on the scale of
$l_B$ ({\it i.e.}, $\xi \gg l_B$) the two functions involved in the integrand of Eq.
(\ref{Step2}) are, in fact, characterized by two very different characteristic
length scales. Indeed, the function
$h_{n_1,n_2}(R,r)$ decays with $R$ on the typical length scale
$R^L_{n}=\sqrt{2n}l_B$ which, for indices $n$ not too big, turns out to be much
smaller  than the characteristic  length scale (of the order of
$\sqrt{(R^L_{n})^2+\xi^2}$) for the spatial variations of the functions
$\Theta_{n,n}\left(R,\omega,\omega\right)$.  Therefore, within the first few Landau
levels
%, the functions $\Theta_{n,n}\left(R,\omega,\omega\right)$ are
%quasi-constant on the cyclotron radius scale, so that 
a good approximation to the integral (\ref{Step2}) is provided by the Laplace's method
owing to the inequality $\xi\gg R_n^L$. Using the formula 
\begin{eqnarray}
\int_{0}^{+ \infty} \!\!\! dx \, I_{0}(ax) e^{-bx^2} = 
\sqrt{\frac{\pi}{4b}} e^{\frac{a^2}{8b}} I_{0}\left(\frac{a^2}{8b} \right),
\end{eqnarray}
we obtain the approximate analytical expression for the LDoS-LDoS correlations
at identical tip energies ($\omega_1=\omega_2=\omega$) for $\omega$ close to 
the energy $E_n$
\begin{widetext}
\begin{eqnarray}
\nonumber
\chi(r,\omega,\omega) \simeq \frac{(2 \pi l_B^2)^{-2}}{2\pi v^2}
\left\{
 \frac{\sqrt{\pi}\xi}{2\sqrt{2}l_B} e^{-(\omega-E_n)^2/(2 v^2)}
\frac{1}{(n!)^2} \frac{\partial^{2n}}{\partial s_1^n \partial s_2^n} \Bigg[
\frac{\sqrt{B_{s_1,s_2}}}{1-s_1 s_2} I_{0} \left( \frac{r^2}{4 l_{B}^2 B_{s_1,s_2}} \right) 
\,e^{ -\frac{r^2}{4 l_{B}^2 B_{s_1,s_2}}} \Bigg]_{s_1=s_2=0} 
\hspace{-1.1cm}
%\, e^{ -r^2/(4 l_{B}^2 B_{s_1,s_2})} \Bigg]_{s_1=s_2=0} \hspace*{-0.4cm}
- e^{-(\omega-E_n)^2/v^2} \right\}.
\hspace*{-0.3cm}\\
\label{chiapprox}
\end{eqnarray}
\end{widetext}
Consequently the nonmonotonous behavior of the correlations seen in Fig.
\ref{Fig2} is connected with the oscillations of the function $h_{n,n}(0,r)$
within Eq.~(\ref{Step2}), and is fully described analytically by 
formula~(\ref{chiapprox}) in the regime $\xi\gg l_B$.
As seen in Fig.~\ref{compare} for two large values of $\xi/l_B$ and for energies
taken at the second Landau level, the agreement between the numerical evaluation of 
the integral in Eq.~(\ref{Step2}) and the analytical approximation (\ref{chiapprox})
is quite excellent. 
We note also that the correlations disappear (for generic values of the
frequency) in the clean limit $v\to 0$ despite the $v^{-2}$ prefactor due to the 
exponential terms in Eq.~(\ref{chiapprox}), while they become singular precisely at 
the Landau level frequency.

\begin{figure}[ht]
\includegraphics{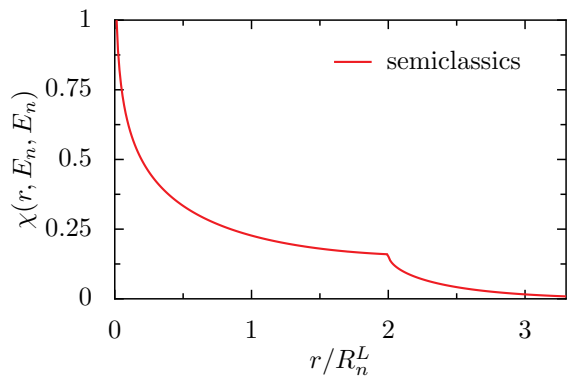}
\caption{(color online) 
LDoS-LDoS correlations at equal energies taken at the Landau level 
($\omega_1=\omega_2=E_n$) as a function of the tips distance $r$ (in units of
$R^L_n$) in the semiclassical approximation ($l_B\to0$) to integral~(\ref{rhorhofinal}).}
\label{ChiSemi}
\end{figure}

We would like to comment here on the semiclassical limit of fixed $R^L_n$ with
$n\rightarrow\infty$ (and vanishing $l_B$) for the LDoS-LDoS correlations. 
In that case, we replace the kernel $F_n({\bf R})$ by 
$\frac{1}{2\pi R^L_n}\delta(|{\bf R}|-R_n^L)$ in Eq.~(\ref{rhorhofinal}),
and obtain numerically Fig.~\ref{ChiSemi}. Not surprisingly, the
quantum oscillations at $r<2R_n^L$ disappear, yet a kink subsists for the
exactly tangent cyclotron trajectories at $r=2R_n^L$, while a logarithmically diverging
correlation occurs at $r=0$. These are classical remnants of the quantum effects 
discussed so far, and this robustness can again be expected according to the geometrical 
argument of Fig.~\ref{ring} (now in the classical limit).
These characteristic features can be understood analytically. Indeed, inserting
Eq.~(\ref{halt2}) within formula~(\ref{Step2}), and performing Laplace's method 
on the $R$-dependent integral (this is valid in the case $\xi\gg R^L)$, we find
\begin{equation}
\chi(r,E_n,E_n)\simeq \frac{(2\pi l_B^2)^{-2}}{ 2 \pi v^2} \Bigg(\frac{\xi}{2} 
\int_{0}^{+ \infty} \!\!\!\!\!\!\!\!  dq J_{0}(rq) [J_{0}(R^Lq)]^2 -1\Bigg).
\label{kink}
%= \frac{1}{r} \{\phantom{.}_2F_1[1/2,1/2,1,a(2R^L/r)]\}^2
\end{equation}
%with $a(x)=(1-\sqrt{1-x^2})/2$ and $\phantom{.}_2F_1$ the hypergeometric function.
Expression~(\ref{kink}) is clearly logarithmically divergent at $r=0$,
but continuous at $r=2R^L$ (the situation of tangent cyclotron orbits). 
The spatial derivative of the correlator can be analyzed in the limit $r\to2R^L$, and gives a 
finite derivative for $r<2R^L$ and a diverging one for $r>2R^L$, explaining the 
kink feature seen in Fig.~\ref{ChiSemi}.
The onset of the logarithmic divergence at $r=0$ can be understood starting from 
the quantum expression~(\ref{chiapprox}), which alternatively reads for $r=0$ and
$\omega=E_n$:
\begin{eqnarray}
\nonumber
\chi(0,E_n,E_n) = \frac{(2\pi l_B^2)^{-2}}{2\pi v^2} 
\Bigg[ \frac{\xi}{\sqrt{2} l_B}
\!\!\int_0^{+\infty} \!\!\!\!\!\!\!\!\!\! du \! 
\left[L_n\left(u^2\right)\right]^2 \! e^{-u^2} \!\!-1\! \Bigg]. \hspace*{-0.5cm}\\
\end{eqnarray}
The above equation is again obtained from Laplace's method in the case 
$\xi\gg R_n^L$  to analytically perform the integral over $R$
in Eq.~(\ref{rhorhofinal}), and used the explicit expression~(\ref{Kneq}) for
the kernel $F_n$, instead of the derivative trick performed in Eq.~(\ref{chiapprox}).
Using the asymptotic formula for the Laguerre polynomial in the large $n$ limit as in Ref. \onlinecite{note2}, we get
$\chi(0,E_n,E_n) \propto \log(n)/R_n^L$ with $R_n^L=\sqrt{2n}l_B$. 
We recover the logarithmic divergence for $n\to\infty$ and fixed $R_n^L$, while 
this expression vanishes as $\log(n)/\sqrt{n}$ for $n\to\infty$ and fixed $l_B$, 
in agreement with Fig.~\ref{Fig2}, where a slight decrease with increasing $n$ 
is seen at $r=0$. We thus stress that quantum mechanics (finite $l_B$) always
regularizes the singular classical behavior, and that maximal correlations are, in fact, obtained (at a given magnetic field, hence fixed $l_B$) for the lowest
Landau levels.

\begin{figure}[ht]
\includegraphics{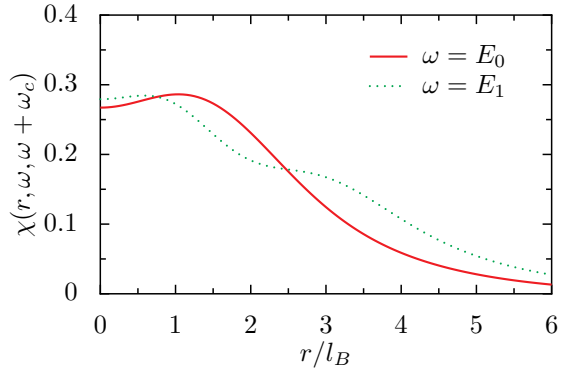}
\caption{(color online) 
LDoS-LDoS correlations at unequal energies ($\omega_1=\omega$ and $\omega_2=\omega+\omega_c$) 
as a function of the tips distance $r$ (in units of $l_B$) for different energies
$\omega$. We used the same parameters as in Fig. \ref{Fig2}.
}
\label{Fig3}
\end{figure}

Let us come back to the quantum case of finite $l_B$ and investigate the
effect of energy detuning in the spatial dependence of the LDoS correlator.
For energies $\omega_1 \neq \omega_2$ a different spatial structure from
the situation of equal energies can be seen in Fig. \ref{Fig3}, which corresponds to the cases
$(\omega_1,\omega_2)=(E_0,E_1)$ and $(\omega_1,\omega_2)=(E_1,E_2)$, the solid
and dotted lines, respectively. Unlike the equal energy cases of Fig.
\ref{Fig2}, the correlations are no more maximally obtained for the tips
distance $r=0$  but for an intermediate tip distance of the order $|R^L_1-R^L_2|$,
as can be guessed again from the geometrical interpretation of Fig.~\ref{ring},
in the case $R^L_1\neq R^L_2$. 
We note that the spatial dependence of the correlations between the first and
second Landau levels is also characterized by an extra mild peak, a reminiscent
feature of the equal energy case.

\subsection{Energy dependence of the correlations}

\begin{figure}[ht]
\includegraphics{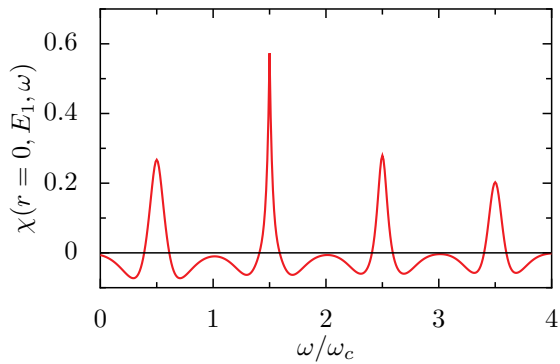}
\caption{(color online) 
LDoS-LDoS correlations for the tip distance $r=0$ and for the energy $\omega_1$
pinned to the first Landau level ($\omega_1=E_1$) as a function of the energy
$\omega_2=\omega$. The other parameters are the same as in Fig. \ref{Fig2}. The
highest peak is obtained for $\omega_2=\omega_1$ and is accompanied by satellite
peaks corresponding to the neighbouring energy levels of the first Landau level. 
}
\label{Fig4}
\end{figure}

We now study the energy dependence of the LDoS-LDoS correlations. We have
represented in Fig. \ref{Fig4} the correlator $\chi(r,\omega_1,\omega_2)$
as a function of the energy $\omega_2=\omega$ for identical tip positions 
({\it i.e.}, for $r=0$) when the first tip energy is pinned to the first Landau level
($\omega_1=E_1$). 
We note the presence of different peaks in the correlations corresponding to 
the sequence $\omega=E_n$ of the Landau levels energy peaks in the local 
density of states. However, it is worth noting that the LDoS-LDoS correlation 
peaks are much sharper than the LDoS peaks because of the sensitive matching of the spatial 
overlaps between the quantum rings associated to the states in Landau level $n_1=1$ and
$n_2=n$. Besides the strongest peak obtained for $\omega=E_1$, a succession of lateral
peaks occurs for all other cyclotron energies $E_n$, showing that nondiagonal
elements of $h_{n_1,n_2}(R,r)$ are also strongly correlated.
More interestingly, the regions surrounding each of these sharp peaks with positive
correlations are characterized by {\it negative} values for the correlator $\chi(0,E_1,\omega_2)$,
thus corresponding to anticorrelations. 
This crossover from positive to negative correlations as a function of the energy 
can be easily understood on the basis of Eq. (\ref{Theta}) at $\xi\gg R^L$, which
is then well approximated by a small-$R$ expansion
\begin{eqnarray}
\nonumber
\Theta_{n_1,n_2}(R,\omega_1,\omega_2) & \simeq & \frac{1}{2 \pi v^2}
\Bigg(
\frac{\xi}{2R}
e^{-\frac{\xi^2}{4R^2} \frac{(\omega_1-\omega_2+E_{n_2}-E_{n_1})^2}{2v^2}}\\
&&\hspace{-2.7cm}
\times e^{-\frac{(\omega_1-E_{n_1})(\omega_2-E_{n_2})}{2v^2}}
-e^{-\frac{(\omega_1-E_{n_1})^2+(\omega_2-E_{n_2})^2}{2v^2}}
\Bigg).
\label{ThetaApprox}
\end{eqnarray}
To obtain this expression, we have developed the disorder correlator
$S_{n_1,n_2}(R)$ [Eq.~(\ref{newcor})] and the linewidth $\Gamma_n$ in the
large $\xi$ limit.
As can be seen in Eq.~(\ref{ThetaApprox}), a competition between two exponential terms 
occurs depending on the relative values of $\omega_1$ and $\omega_2$.
For $\delta\omega\equiv\omega_1-\omega_2+E_{n_2}-E_{n_1}=0$, 
the first term in the right-hand side (r.h.s) of Eq. (\ref{ThetaApprox}),
which is large and positive (of the order of $\xi/R$), dominates over the second
exponential contribution, and one gets positive LDoS-LDoS correlations, as
already noted. For $\delta\omega$ slightly different from zero, the first term
decreases exponentially fast (note the large $\xi^2/R^2$ prefactor within the
exponential), resulting in the sharp peak observed in Fig.~\ref{Fig4} near 
coinciding energies. At increasing $|\delta\omega|$,
the second term in the r.h.s of Eq. (\ref{ThetaApprox}), which has a small and 
negative amplitude, is characterized by a slower exponential decrease, and then gives 
the main contribution to the correlations.
Coming back to the initial Eq. (\ref{chi}) for the correlator
$\chi(r,\omega_1,\omega_2)$, we conclude that the square average 
density dominates quickly at increasing energy detuning.

Analytical insight can be obtained by further assuming the classical 
regime of $R^L\gg l_B$ (hence high Landau levels). Using then
the approximation~(\ref{halt2}) within formula~(\ref{Step2}) together with 
Eq.~(\ref{ThetaApprox}) (which is valid in the limit $\xi\gg R^L$) we get
the semiclassical approximation for the LDoS correlations at coinciding tip
position $r=0$
\begin{widetext}
\begin{eqnarray}
\nonumber
\chi(0,\omega_1,\omega_2) &=& \frac{\xi}{2\pi v^2 (2\pi l_B^2)^2} 
\sum_{n_1,n_2}
\int_0^{+\infty}\!\!\!\!\!\!\!\!\! dq \, q J_0(R_{n_1}^L q) J_0(R_{n_2}^L q) 
\int_0^{+\infty}\!\!\!\!\!\!\!\!\! dR \, J_0(Rq) 
e^{-\frac{(\delta\omega)^2}{2v^2}\frac{\xi^2}{4R^2}} 
e^{-\frac{(\omega_1-E_{n_1})(\omega_2-E_{n_2})}{2v^2}} \\
&& -\big<\rho(0,\omega_1)\big>\big<\rho(0,\omega_2)\big>.
\label{chiaprox2}
\end{eqnarray}
\end{widetext}
Here the double sum over $(n_1,n_2)$ is again constrained by the
external frequencies $(\omega_1,\omega_2)$, and reduces to a single
term in case of sharply defined LDoS-LDoS correlation peaks,
giving rise to a single peak at $\omega_1-\omega_2=E_{n_1}-E_{n_2}$.
Let us focus first on the case where $n_1=n_2$, so that
$R_{n_1}^L=R_{n_2}^L=R^L$.
We see that the above integrand in Eq.~(\ref{chiaprox2}) then behaves as 
$1/q$ for vanishing $\delta\omega$, because $|J_0(R^L q)|^2$ looses its oscillatory
character at large momentum $q$. One obtains thus a logarithmically diverging peak 
$\chi(0,\omega_1,\omega_2)\sim \log|v/\delta\omega|$, related to the $\log|l_B/r|$ 
spatial divergence found previously in the semiclassical limit at small
intertip distance.
However, for nonzero $n\equiv n_1-n_2$, the LDoS-LDoS correlation peak at
$\omega_1-\omega_2=E_n$ is no more logarithmically diverging when
$\delta\omega\to 0$. This is because
the product $J_0(R_{n_1}^L q) J_0(R_{n_2}^L q)$ behaves as $\cos(l_B^2 n q/R^L)/q$
for large momentum $q$, as can be seen using the expression for the cyclotron radii 
$R_{n_{1,2}}^L = R^L \pm \frac{l_B^2 n}{2 R^L}$ in the limit $n\ll n_1,n_2$,
where $R^L\equiv \sqrt{n_1+n_2}l_B$.
In order to capture the relevant energy scale at small $\delta \omega$, we can
make the change in variables $k=(\xi |\delta\omega|/v) q$ and 
$z=(v/\xi \Omega_n)R$ with $\Omega_n = n v l_B^2/(R^L\xi)$ 
into Eq.~(\ref{chiaprox2}), and use the asymptotic form of the Bessel functions
for $|\delta\omega|\ll \xi v/R^L$. After integration over $k$, this provides the 
deviation of the LDoS correlations from the $n^\mathrm{th}$ peak value 
(at $\delta\omega=0$)
\begin{equation}
\delta \chi = \frac{(2\pi l_B^2)^{-2}}{2\pi^2 v^2} 
\frac{\xi}{ R^L} 
 \int_1^{+\infty} \!\!\!\! dz \; 
\frac{e^{-\left|\frac{\delta\omega}{ \Omega_n}\right|^2
\frac{1}{8z^2}}-1}{\sqrt{z^2-1}}.
\end{equation}
The new energy scale $\Omega_n$ therefore sets the width of the 
$n^\mathrm{th}$ correlation peak. Since $\Omega_n\ll v$ in the regime $\xi\gg R^L\gg l_B$, 
we recover the fact that the LDoS-LDoS correlations are more sharply 
defined than the average LDoS peaks. The linear increase of $\Omega_n$ with 
$n$ also explains the progressive smearing of the correlations at increasing 
energy detuning of the tips (see Fig.~\ref{Fig4}).

We now discuss the situation of strongly overlapping Landau level,
$\omega_c \ll v = \sqrt{\left<V^2\right>}$, so that the average density of 
states becomes structureless. In that case, the sharper peaks in the LDoS-LDoS 
correlations can survive for smooth disorder ($\xi\gg l_B$), under the condition 
$v l_B/\xi \ll \omega_c$ for the lowest Landau levels. 
A similar result was already found by Rudin {\it et al.} \cite{Rudin}, 
who studied the energy dependence of the {\it local} LDoS-LDoS correlations in a 
{\it weak} magnetic field and long classical orbits $R^L\gg \xi$ in the
diffusive regime. The disorder dependence of correlator~(\ref{chiapprox}) in
Ref.~\onlinecite{Rudin} was also in $1/v^2$ as in the present paper.

It is worth stressing that the condition to obtain sharp peaks in the
LDoS-LDoS correlations corresponds 
precisely to the absence of local Landau level mixing. 
Indeed, transitions between adjacent Landau levels provide~\cite{Champel2008} a typical 
energy scale $l_B^2 | {\bm \nabla} V|^2/\omega_c$, so that 
our calculation is controlled when the parameter $(l_B^2/\xi^2)(v^2/\omega_c^2)$ is small.
Clearly, large overlap in the 
average DoS can be compatible with no mixing, because this quantity relates to 
global properties (long-wavelength fluctuations) of the smooth disorder. In contrast, 
only the correlations of the LDoS can feel the local interplay of disorder 
and Landau quantization, and reveal whether the Landau index $n$ stays a good quantum 
number or not.

Interestingly, a logarithmic singularity for the $n=0$ peak
and an energy width $\Omega_n$ proportional to $n^2$ for the $n^\mathrm{th}$
peak were obtained in the semiclassical diffusive regime~\cite{Rudin}. This is only 
slightly different from our results in the semiclassical limit, showing the continuity 
of the present physics from low to high magnetic fields. We emphasize again that
quantum effects at finite magnetic length regularize the spurious divergences
generically related to the semiclassical approximation.

Finally, we note that some of the energy-related features on the LDoS correlations 
discussed above have already been reported experimentally~\cite{Konemann,Holder}
in heavily doped three-dimensional (3D) GaAs semiconductors.
For instance, conductance anticorrelations were observed~\cite{Konemann} 
in the presence of Landau levels, while the narrowing of the LDoS fluctuations 
has been found~\cite{Holder} at weak magnetic field. However, these experimental studies 
were performed using resonant tunneling impurity, hence at a fixed position, asking 
for a different analysis from what was performed here (besides the 3D character of 
the studied samples).
Indeed, averaging was performed either over the applied voltage~\cite{Holder} 
or magnetic field~\cite{Konemann}, and in the later case, the Landau levels are
simply lost. Spatial averaging in a STM configuration, as proposed in our work, 
should allow a greater control of the LDoS correlations (in terms of the applied
magnetic field and tip voltages). Moreover, this offers a way to investigate 
spatial correlations of the LDoS, that could not be assessed with previous 
experimental techniques.

\section{Conclusion}
We have studied theoretically the two-point correlations of the local density of 
states in a disordered two-dimensional electron gas under a large magnetic field. 
A rich spatial dependence of the correlations was found, which can be
qualitatively explained by geometrical overlaps of the two quantum cyclotron rings
that roughly describe circular wave functions.
The energy behavior of the correlations was shown to provide sharp peaks when
the frequency detuning matches integer multiple of the cyclotron frequency,
similar to the low magnetic field results of Rudin {\it et al.} \cite{Rudin}.
These sharp peaks in the LDoS correlations reveal that Landau levels correspond 
to well-defined quantum numbers, information that cannot be easily gathered from
the average density of states only, where large overlaps are usually reported
experimentally~\cite{Hashi2008} even at large magnetic fields.
We have also emphasized here that LDoS correlations can be either positive or
negative, depending on the degree of frequency mismatch.
We would also like to mention that at energies strictly coinciding with
the centers of Landau levels (quantum Hall critical point), the LDoS
correlations exhibit a power-law behavior at long distance, reflecting
the multifractality of critical wave functions~\cite{Chalker,Mirlin}. This more
complex behavior is beyond the scope of the present paper, which did not consider quantum tunneling between closed drift orbits. Such effects are, however, only relevant at very low temperatures for a smooth potential.

We end up by noting that recent experimental progresses for electron gases
confined at the surface of InSb semiconductor~\cite{Hashi2008} and also in
graphene~\cite{Miller2009,Miller2010} allow high spatial and energy resolution 
measurements of the local density of states. The disorder averaged LDoS and its 
correlations can, in principle, be straightforwardly obtained from the
experimental data by averaging over large scale spatial maps. Because both
the width of the disorder distribution and the correlation length of the 
random landscape can be extracted from the knowledge of the average LDoS, our
predictions could be tested even quantitatively without extra fitting parameter.
In the case of graphene, extension of the present work can be
straightforwardly done following the results of Ref.~\onlinecite{Champel2010}.
The spinorial form of the wavefunction is expected to lead to more complex
spatial structures, yet with similar general behavior to that discussed here.

\begin{acknowledgments}
We are thankful for the hospitality of APCTP at POSTECH, where this work
was initiated. M. E. R. acknowledges the support of DOE Grant No. 
DE-FG02-06ER46313.
\end{acknowledgments}


\begin{thebibliography}{99}
\bibitem{note1}
The equivalence between Eqs. (\ref{Kn}) and (\ref{Kneq}) can be established
straightforwardly using the definition of the Laguerre polynomials in terms of
the generating function $g(x,z)=\exp\left[-xz/(1-z)\right]/(1-z)$.
\end{thebibliography}
\end{document}